\documentclass[twocolumn,
showpacs,
 amsmath,amssymb,amsfonts,
 aps,
prl,
]{revtex4-1}

\usepackage{subfigure}%
\usepackage{graphicx}
\usepackage{dcolumn}
\usepackage{bm}

\usepackage{xcolor}

\begin{document}

\preprint{APS/123-QED}

\title{ A comparative study of crumpling and folding of thin sheets  }

\author{S. Deboeuf$^{1}$, E. Katzav$^2$, A. Boudaoud$^3$, D. Bonn$^4$, M. Adda-Bedia$^5$}  
\affiliation{$^1$Universit\'e Paris-Est, Laboratoire Navier (UMR 8205), CNRS, ENPC, IFSTTAR, F-77420 Marne-la-Vall\'ee \\
$^2$Department of Mathematics, King's College London, Strand, London WC2R 2LS, UK \\
$^3$RDP, ENS Lyon, 46 all\'ee d'Italie, 69007 Lyon, France \\
$^4$Institute of Physics, University of Amsterdam, Science Park 904, Amsterdam, the Netherlands \\
$^5$Laboratoire de Physique Statistique, Ecole Normale Sup\'erieure, UPMC Paris 6, Universit\'e Paris Diderot, CNRS, 24 rue Lhomond, 75005 Paris,  France}

\date{\today}

\begin{abstract}
Crumpling and folding of paper are at first sight very different ways of confining thin sheets in a small volume: the former one is random and stochastic whereas the latest one is regular and deterministic.  Nevertheless, certain similarities exist. Crumpling is surprisingly inefficient: a typical crumpled paper ball in a waste-bin consists of as much as $80\%$ air. Similarly, if one folds a sheet of paper repeatedly in two, the necessary force  becomes so large that it is impossible to fold it more than 6 or 7 times. Here we show that the stiffness that builds up in the two processes is of the same nature, and therefore simple folding models allow to capture also the main features of crumpling.  An original geometrical approach shows that crumpling is hierarchical, just as the repeated folding. For both processes the number of layers increases with the degree of compaction. We find that for both processes the crumpling force increases as a power law with the number of folded layers, and that the dimensionality of the compaction process (crumpling or folding) controls the exponent of the scaling law between the force and the compaction ratio.
\end{abstract}

\maketitle

\begin{figure}[htb]
\centering
\hspace*{20mm} (a) \hfill (b) \hspace{15mm} (c) \hspace*{7mm} \\
 {\includegraphics[keepaspectratio=false,height=0.15\linewidth]{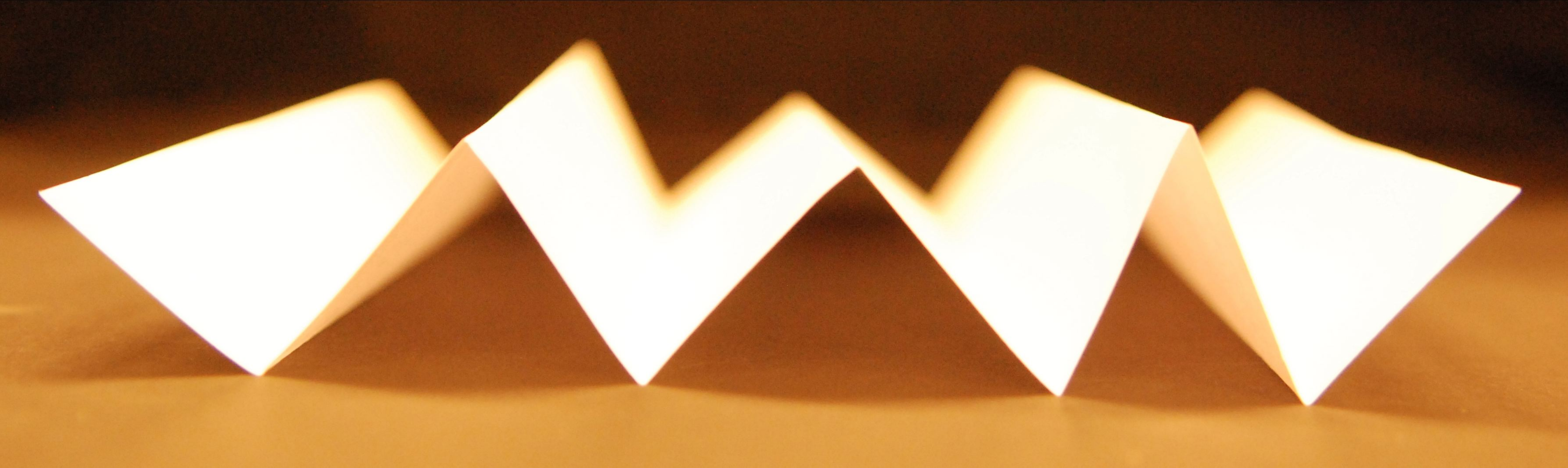}}
 {\includegraphics[height=0.15\linewidth]{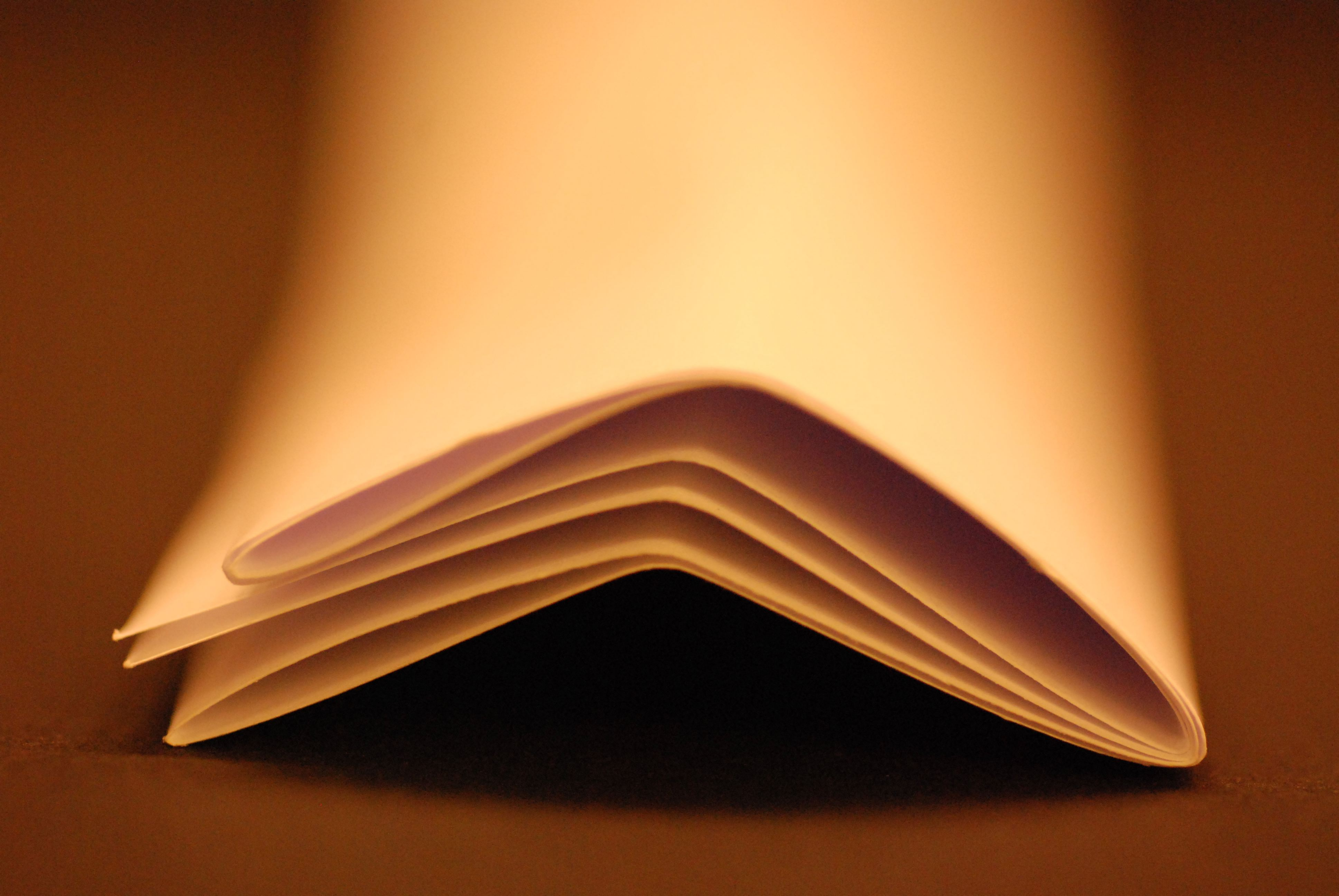}}
 { \includegraphics[height=0.15\linewidth]{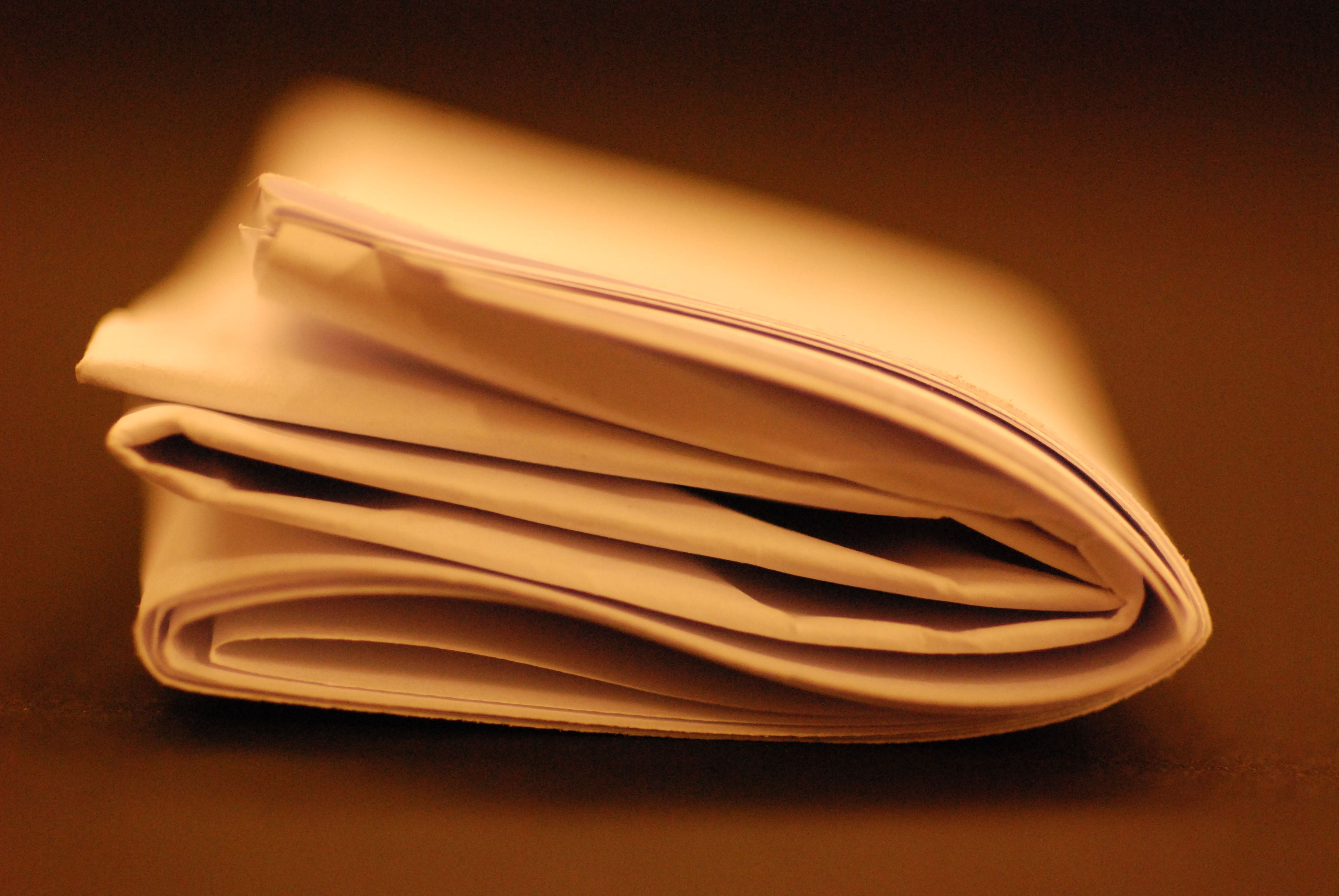}}
\caption{Hierarchical folding of a sheet in different dimensionalities. (a) a 1d-like sheet folded in 1d; (b) a 1d-like sheet folded in 2d; (c) a 2d-like sheet folded in 3d. The three types of folding processes are referred to as 1d, 2d and 3d compactions.}
    \label{fig0}
\end{figure}

It is easy to verify that the maximum number of times one can {\sl fold} a sheet of paper is only $6$ or $7$, which surprisingly is  independent of the initial size of the sheet. Quantitatively, elasticity theory allows to write the relation between the compaction force and the number of times one can repeatedly fold a piece of paper in two. This follows from the scaling of the bending rigidity $B$ with the thickness $h$ of the folded sheet~\cite{landaulifshitz},  $B = E h^3/12(1-\nu^2) $, where $E$ is the Young modulus and $\nu$ the Poisson ratio. For a sheet of initial size $D_1 \times D_2$ folded along the direction  $D_2$, the compression energy, $E_c$, injected in the system  should be compared to the typical energy dissipated in the fold. One writes $E_c=FD_2$, where $F$ is a characteristic compression force applied along the direction $D_1$. 
Since most of the folded sheet remains flat and the region which is irreversibly deformed  
 is straight along $D_2$ (i.e. its gaussian  curvature equals to zero ; See Fig.~\ref{fig0}), the energy dissipated in the fold is estimated from the elastic bending energy, $E_{el}$, concentrated in a region of length $D_2$ and width $h$ with a curvature $1/h$~\cite{landaulifshitz}; this leads to $E_{el}\sim B h D_2/h^2=BD_2/h$~\cite{Dias12}. The balance of these two energies leads to $F_0\sim B/h $ for   the elementary force needed to create a unique fold. When the sheet is  folded $n$ times repeatedly leading to the hierarchical creation of folds, its effective thickness  and bending rigidity become $h_n \rightarrow 2^{n} h$ and $B_n \rightarrow 2^{3n} B$ assuming no slip between layers (this hypothesis becomes increasingly consistent for large $n$). Consequently, for a sheet folded $n$ times, the energy balance gives
\begin{equation}
F(n) \sim B_n/h_n \sim F_0 2^{2n} \,.
 \label{FN}
\end{equation} 
 Thus the force is independent of the initial size of the sheet and grows exponentially with the number of folding events $n$. The exponential dependence is the reason why one cannot fold a sheet indefinitely by hand or by applying a finite force; the elementary force $F_0$ is estimated with typical values $E = 10^9$Pa, $h=10^{-4}$m leading to $F_0 \approx 1$\,N. Then,  for $n = 6$, $F$ becomes of the order of kiloNewtons,  which is larger than the maximal force any person can exert and then sets a limit on the achievable number of successive foldings. 

Repeated folding in two is  not the only possible way to fold; here we consider three basic regular processes (Fig.~\ref{fig0}), which are   prototypical foldings in various dimensionalities. In cases (a) and (b), the sheet is thought of as a 1d-like sheet, since it is folded along one direction only. However compaction of case (a) is not isotropic, contrary to case (b). The latter is seen as isotropic compaction within a disk, whereas the former is seen as unidirectional compaction inside an elongated rectangle such as in~\cite{Roman02}. Finally, for case (c) the sheet is truly two-dimensional and compacted in a sphere. The number of folded layers $N=h_n/h$ after $n$ folding events and the related compaction ratio $\phi$ (defined as $\phi \equiv D/\Delta$, where $D$ and $\Delta$ are the initial and final size of the sheet) then depends on the precise geometry and dimensionality of the compaction process. Folding a $d$-dimensional sheet in $(d+1)$-dimensional space, $N=2^n$, whereas for the $1d$--$1d$ case (a), $N=n$. Also, for foldings (a) and (b), $N=\phi$, whereas in the $3d$-like folding (c), $N=\phi^2$. Using similar arguments leading to Eq.~(\ref{FN}), one finds a generic power law relation between the force $F$, the compaction ratio $\phi$ and the number of folded layers $N$:
\begin{eqnarray}
&\mbox{1d compaction, case (a)~: }&
F(N) = F_0 N \sim F_0 \phi 
 \label{FN1}\\
&\mbox{2d compaction, case (b)~: }&
F(N) = F_0 N^2 \sim F_0 \phi^2
 \label{FN2}\\
&\mbox{3d compaction, case (c)~: }& 
F(N) = F_0 N^2 \sim F_0 \phi^4 
 \label{FN3}
\end{eqnarray}
It is important to note that we described folds as the result of an irreversible process occurring in a small region of size $h$ and characterized by a zero gaussian curvature. Consequently, energy scalings are {\it different} from those obtained  for singular ridges~\cite{witten07,Roman12} that are reversible. Moreover, our estimate for the  energy dissipated in the fold should be taken as a lower bound  because the contribution of vertices~\cite{BenAmar97} and other possible length scales associated with plastic events are  neglected. Our approach is inspired by crack modelling in the framework of linear elastic fracture mechanics where the process zone near the crack tip is neglected and the dissipation is estimated from its balance with the far-field elastic energy.
 
\begin{figure}[htb]
\centering
\hspace*{17mm} (a) \hfill (b) \hspace*{20mm} \\
  {\includegraphics[height=0.34\linewidth]{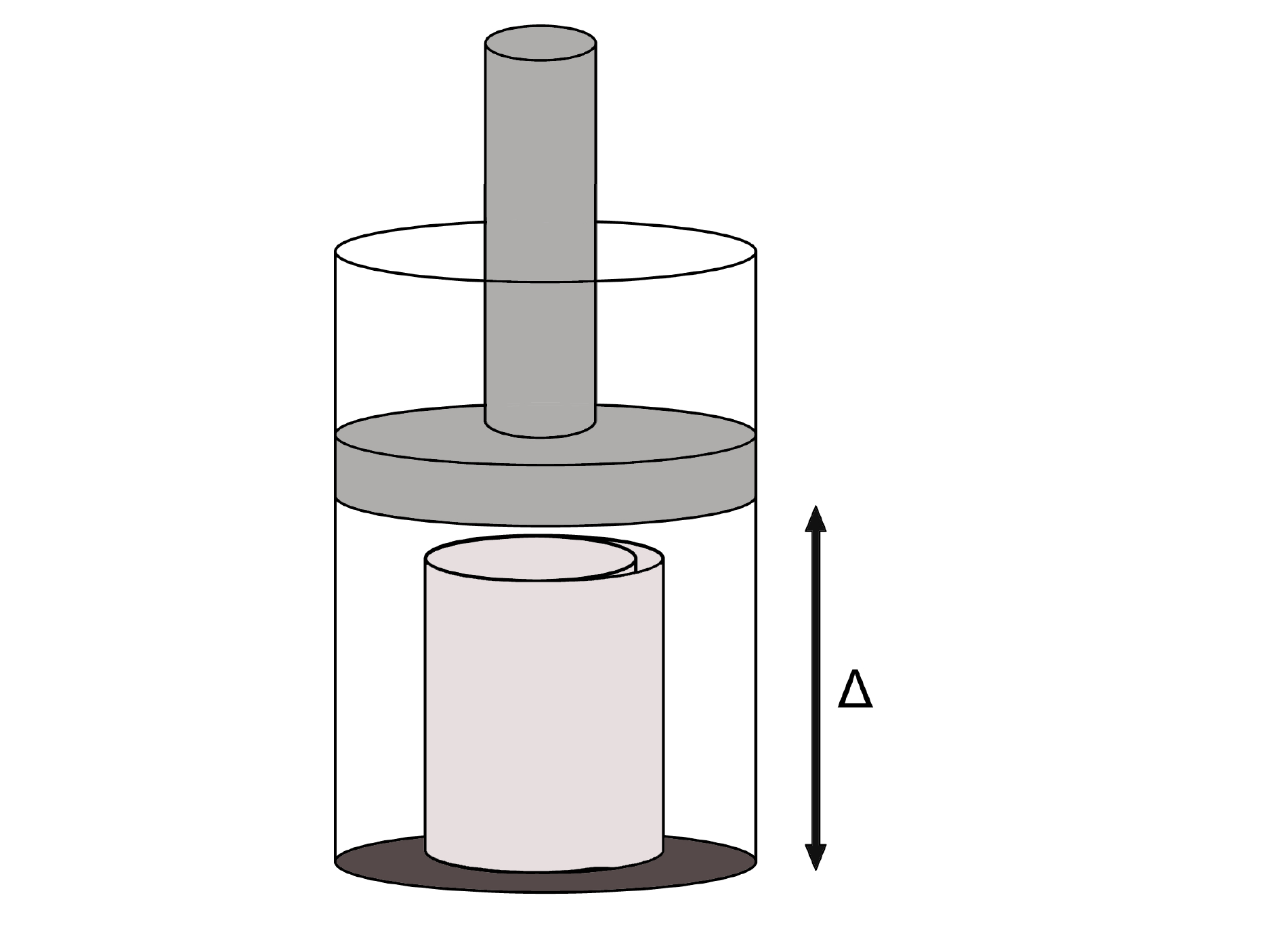}}
 {\includegraphics[width=0.48\linewidth]{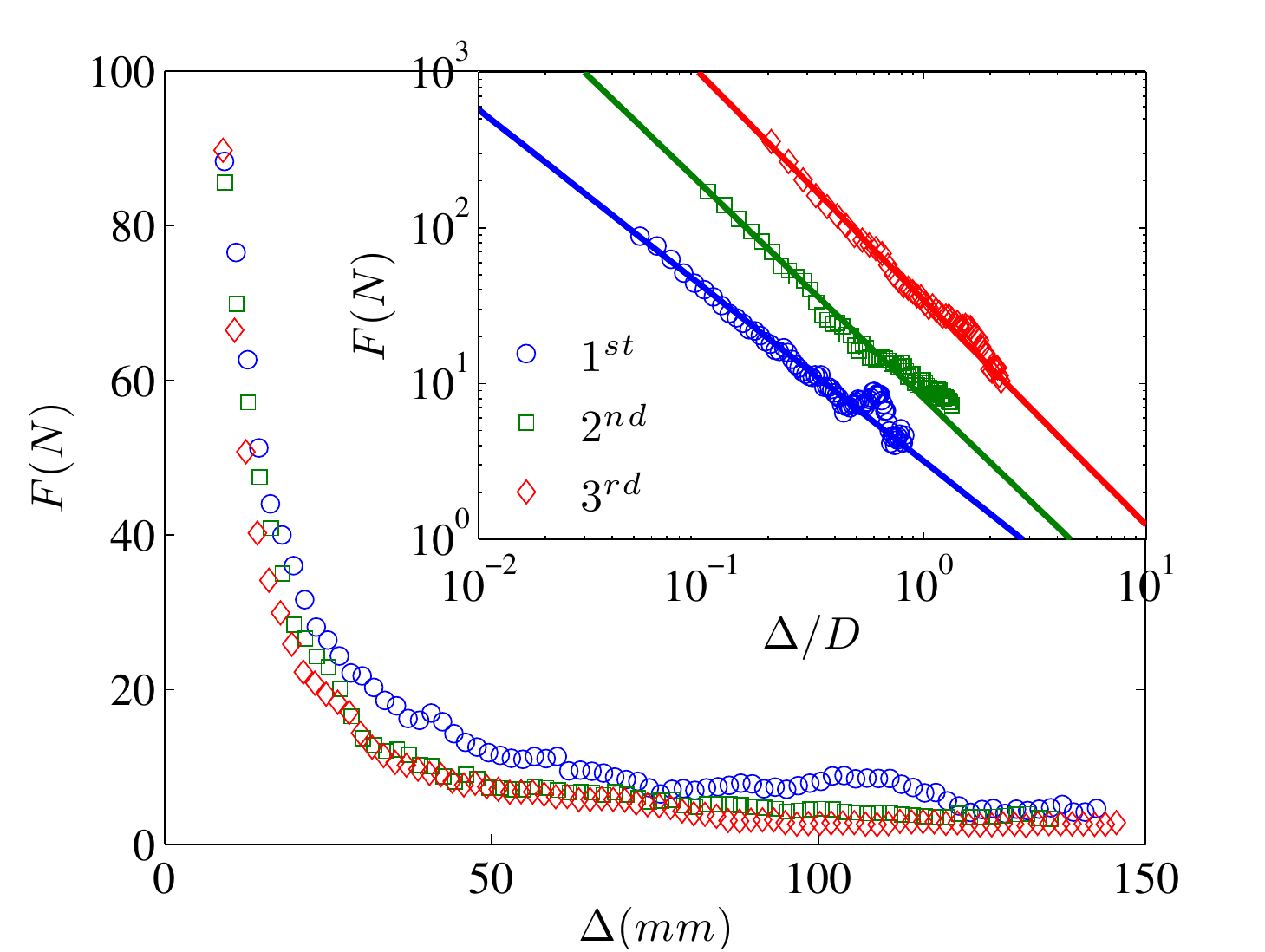}}\\
\hspace*{17mm} (c) \hfill (d) \hspace*{20mm} \\
  { \includegraphics[width=0.48\linewidth]{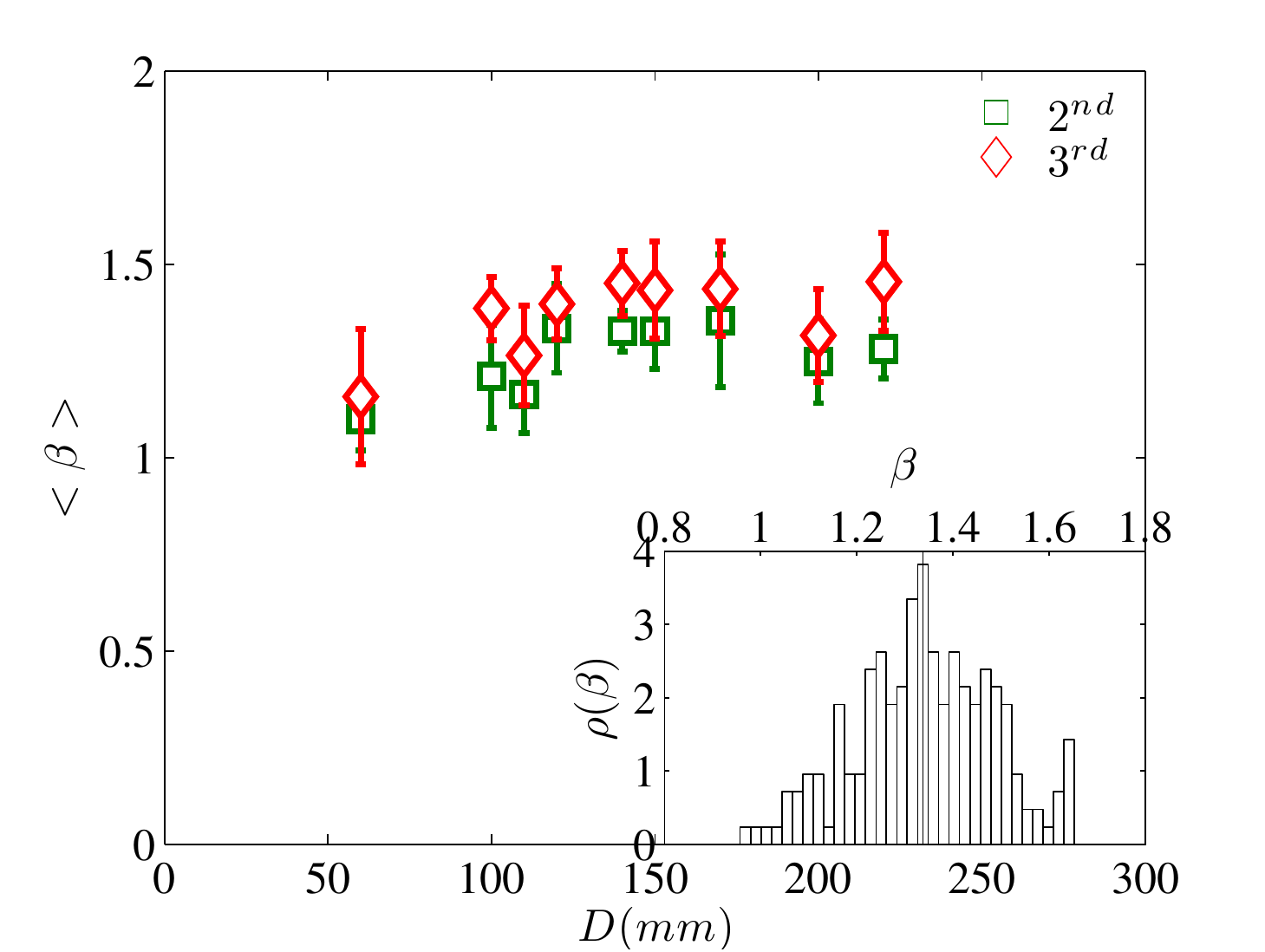}}
  { \includegraphics[width=0.48\linewidth]{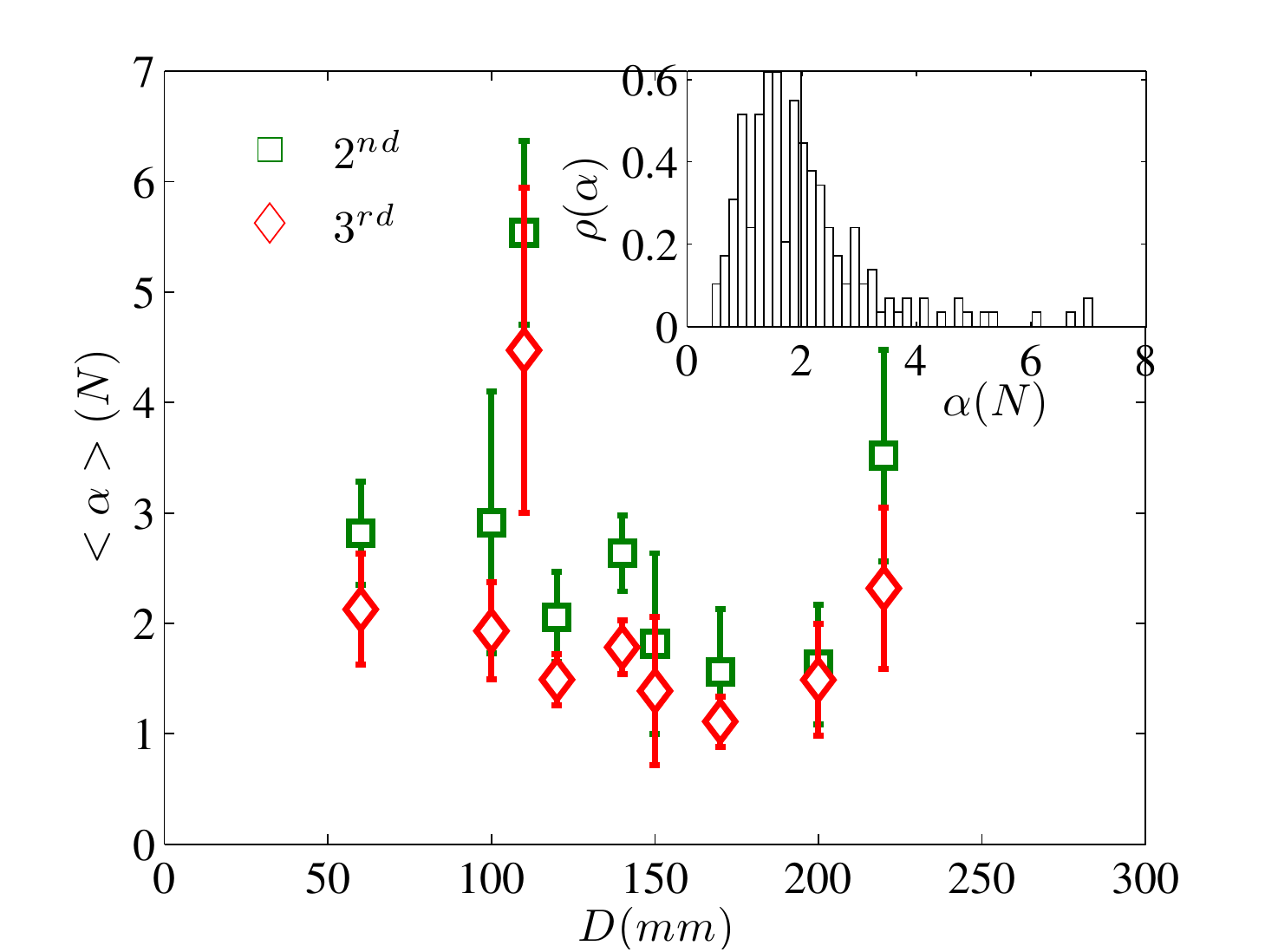}}
    \caption{(a) Setup used to measure the force $F(\Delta)$ during crumpling. (b) Typical force-distance  $F(\Delta)$ and force-compaction ratio  $F(\Delta/D)$  curves on a log-log scale and shifted with respect to each other for clarity in inset, with Kraft paper of size $D= 15$cm in a cell of diameter $6$cm, for the $1^{st}$ (blue circles), $2^{nd}$ (green squares) and $3^{rd}$ (red diamonds) crumpling rounds. The lines are the fits to a power law (Eq.~\ref{EqnExp}): ($\alpha$, $\beta$) are ($3.74$\,N, $1.1$), ($1.90$\,N, $1.4$) and ($1.43$\,N, $1.5$) for the $1^{st}$,  $2^{nd}$ and $3^{rd}$ crumpling rounds respectively. (c) Exponent $\beta$ of the power law fit as a function of the paper size $D$ and its probability distribution function in inset 
 for the  $2^{nd}$ (green) and $3^{rd}$ (red) crumpling rounds. (d) Characteristic force scale  $\alpha$ of the power law crumpling force as a function of  $D$ and its probability distribution function in inset. For (c) and (d) many experimental realizations were averaged for constant values of $D$.}
    \label{fig1}
\end{figure}

One may wonder to which extent our models for regular folding describe our crumpling measurements and whether crumpling process shall  be viewed as arising from successive folding events.
To compare the regular folding with the {\it crumpling} of paper, we first show experimentally that here  the force also increases as a power law with the degree of compaction, with exponents in accordance with predictions dependant on the dimensionalities of the compaction process. For this purpose, a sheet of paper of characteristic size $D$, is placed into a rigid cylindrical cell, in which a piston connected to a force transducer compacts the sheet at constant speed (Fig.~\ref{fig1}a). The materials used here (Kraft and regular printing papers) have been chosen for their low ductility.  The experimental force-distance curves show a very strong increase of the force upon compaction (Fig.~\ref{fig1}b), for both 'virgin' sheets (crumpled for the first time) and 'trained' sheets (crumpled for the second or third time). For the latter the force-compaction ratio curves turn out to be independent of the initial preparation of the sheet inside the cylinder within the experimental accuracy. Then, the measured curves for different types of paper of similar properties (thickness $h=10 \mu m$ and Young modulus $E\simeq10^9Pa$), different sheet sizes, cells and compaction speeds are all described by a power law:
\begin{equation}
F(\Delta) = \alpha \left( \frac{\Delta}{D} \right)^{-\beta}\, = \alpha \phi^{\beta}\, ,
 \label{EqnExp}
\end{equation}
where $\alpha$ is a characteristic force scale and $\Delta$ is the gap between the piston and the bottom of the cell (Fig.~\ref{fig1}a). While the data range for $\Delta$ is small, this behaviour is robust over all the 150 realizations. The statistical $\chi^2$ test for goodness-of-fit confirms the relevance of the power law in comparison with other fits. The exponent $\beta$ of the power law divergence is  $\beta \approx 1.3 $ (Fig.~\ref{fig1}c), a value between $1$ and $2$, those expected for {\it ordered} folding in 1d and 2d. We argue below that this is due to the anisotropy of the compaction process in our experiment. Effectively, compaction here is quasi $1d$, since loading is applied mainly in one direction. However, the setup also allows for compaction in the perpendicular direction, which would rather be a $2d$ process. Moreover, the characteristic force scale $\alpha$ is independent of size $D$: $\alpha \approx 2$\,N (Fig.~\ref{fig1}d) which is of the same order of magnitude as the characteristic force, $ F_0 \approx 1$\,N, calculated for the folding.

\begin{figure}[htb]
\centering
\hspace*{20mm} (a) \hfill (b) \hspace*{25mm} \\
{\includegraphics[angle=90,height= 0.35\linewidth]{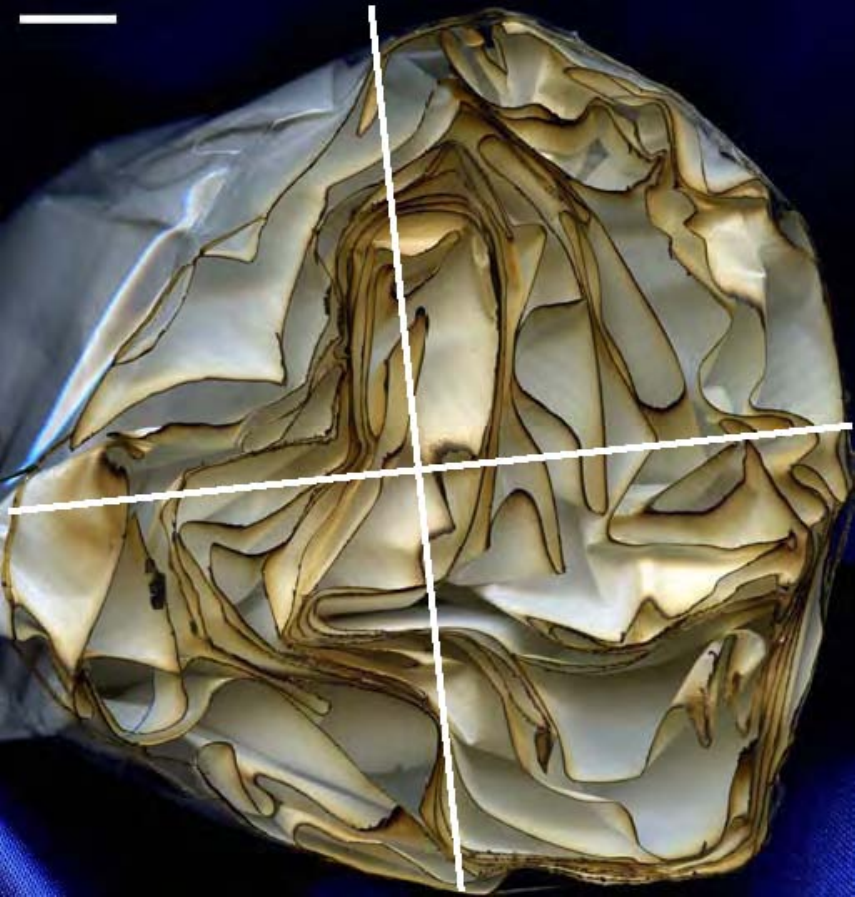}}
{\includegraphics[height = 0.35\linewidth]{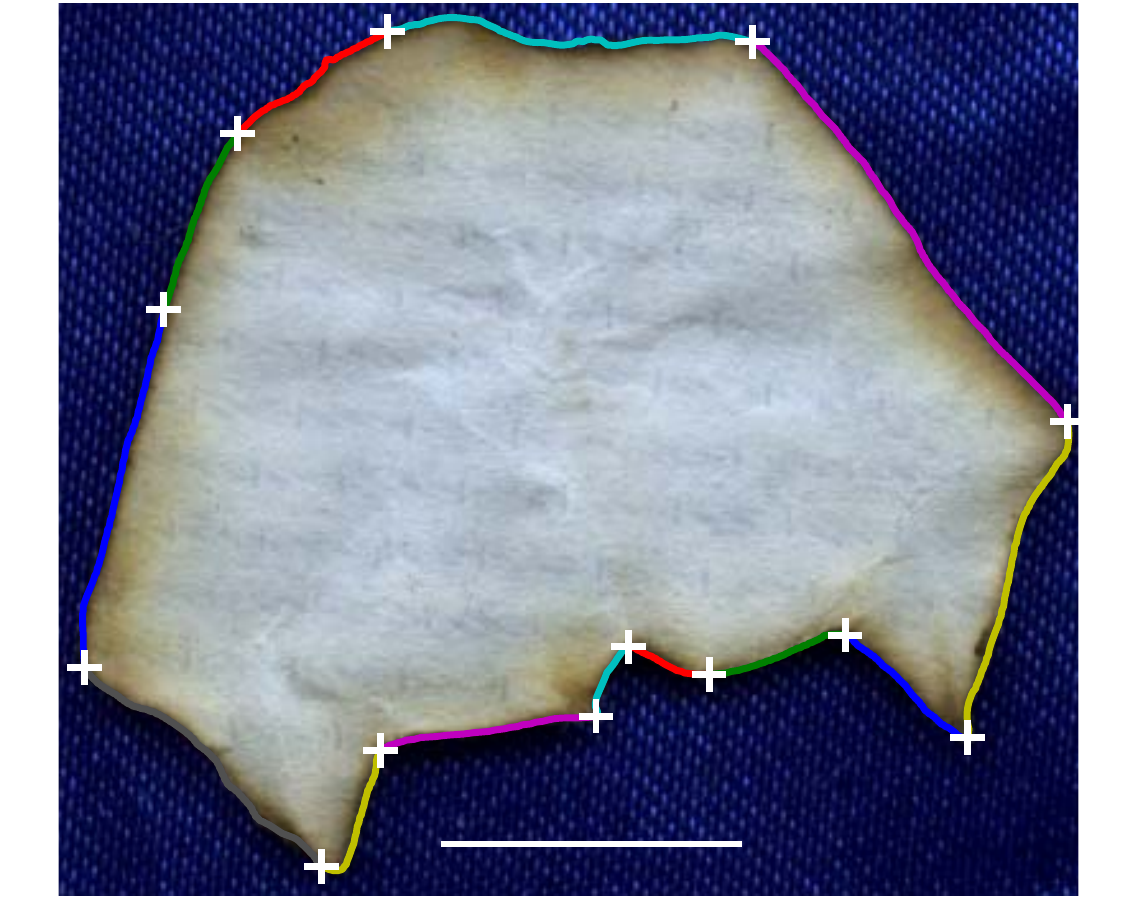}} \\
\hspace*{20mm} (c) \hfill (d) \hspace*{20mm} \\
{\includegraphics[width = 0.48 \linewidth]{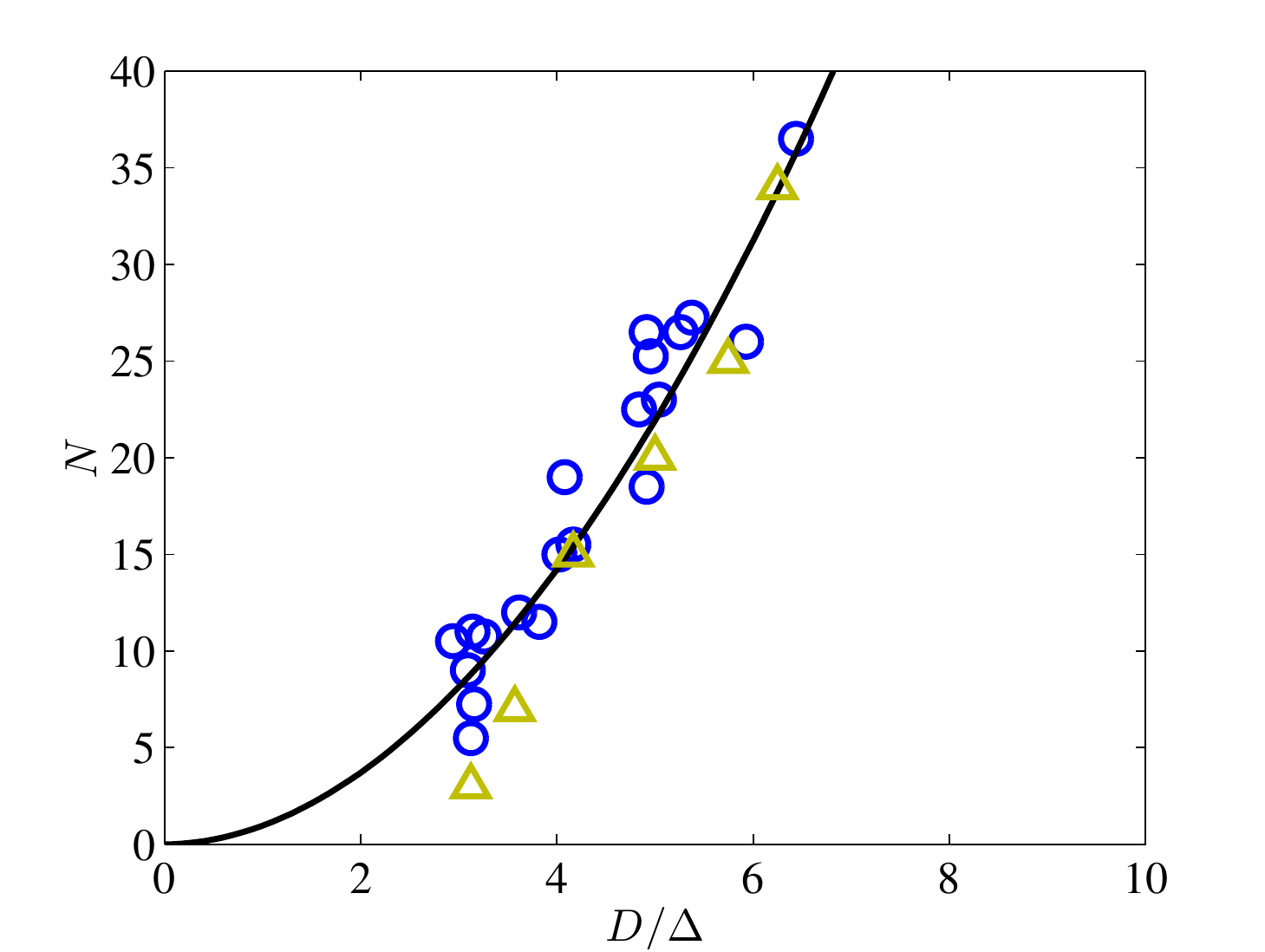}}
{\includegraphics[width= 0.48\linewidth]{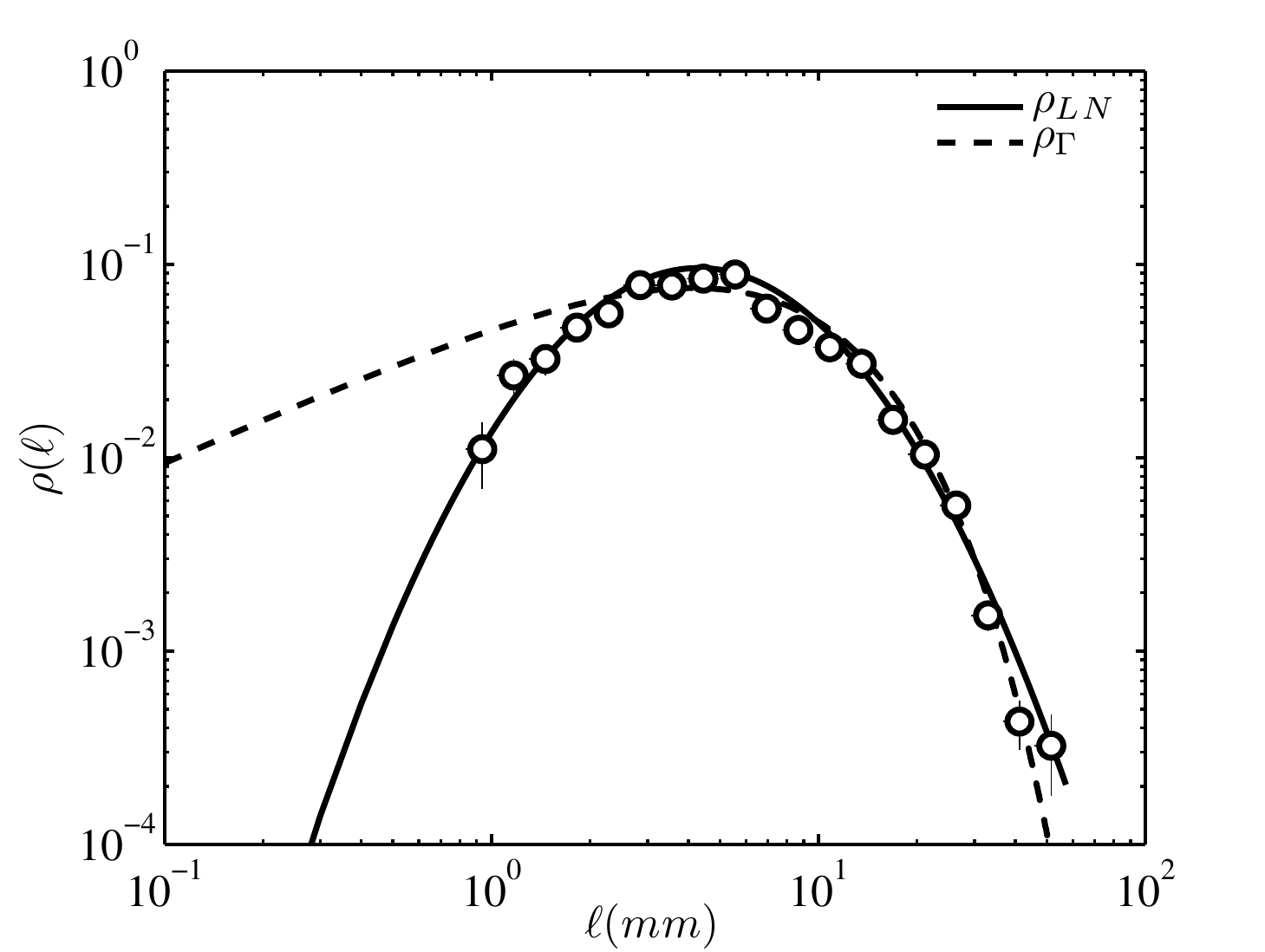}}
\caption{(a) Picture of a crumpled cross-section and two orthogonal directions used to extract the number of folded layers $N$. (b) Picture of a piece of an uncrumpled cross-section and its segmented edge. Scale bars are $10$mm.  (c)  Number of folded layers $N$ as a function of compaction ratio $D/\Delta$ for crumpled balls of Kraft paper (circles) and regular printing paper (triangles). The line is the curve $N=(D/\Delta)^2$.  (d) Probability density function $\rho(\ell)$ of the length of segments on a log-log scale for a Kraft paper sheet crumpled in a ball of diameter $\Delta\approx 80$mm and
compaction ratio $\phi\approx6$. The continuous and dashed lines are the log-normal and  Gamma distributions respectively, with the same mean and variance as the data. The error bars $\delta\ell$ and $\delta\rho$ are the bin width $\delta\ell$ and the standard deviation $\delta\rho=\rho/\sqrt{n}$ of the histograms $n(\ell)$.  }
 \label{fig2}
\end{figure}

A second step towards understanding the analogy between crumpling and folding is to establish the relation between the degree of compaction and the number of folds for the experimental crumpled configurations. To achieve this, we characterize the geometry of the crumpled paper, using an original approach that makes use of the properties of folds and facets in cross-sections of crumpled samples. Sheets of different paper types and sizes $D$ are crumpled into hand-made balls at different degrees of compaction. A cross-section is obtained by cutting the crumpled ball in two with a slowly moving hot wire \cite{deboeuf09}. The overall size of the crumpled configuration, $\Delta$, is defined as the largest diameter of the resulting cross-sectional area. In this cross-section, the number of paper layers is measured in two orthogonal directions passing through the center (Fig.~\ref{fig2}a), and averaged to obtain the mean number of folded layers $N$ in the crumpled configuration. By this method, we ensure that $N$ is defined as in the folding model introduced above ($N=h_n/h$). The number of folded layers $N$ is described by a power two dependance on the degree of compaction $N = (D/\Delta)^2=\phi^2$, with a prefactor equal to 1 (Fig.~\ref{fig2}c). This important result is exactly the same as that observed for 3d folding (Fig.~\ref{fig0}c) showing that the geometry of folding and crumpling is the same.

A possible difference between the two situations is that while the repeated folding is a hierarchical process, this is not clear for crumpling. To investigate whether the crumpling is also hierarchical, we characterize the lengths of folds and facets in cross-sections of crumpled samples. For this purpose, the cut crumpled sheet is reopened carefully and thel uncrumpled pieces, with possibly several holes, are scanned (Fig.~\ref{fig2}b). The edges of their boundaries and holes are detected automatically and broken down into segments delimited by kinks~\cite{sultan06}, by using a `Split and Merge' algorithm~\cite{pavlidis74} for the segmentation. The planar two-dimensional cross-section of the crumpled sheet bears information on the full three-dimensional crumpled configuration: the ensemble of segments samples the facets delimited by folds, so that one segment length is related to the characteristic size of the facet or equivalently to the characteristic distance between folds. We use this to asses the nature of the crumpling. To do so, we compare the distribution of lengths $\rho(\ell)$ with a log-normal distribution and a Gamma distribution; the former characterizes a hierarchical process~\cite{Wood02,sultan06,witten07}, whereas the latter accounts for random processes~\cite{Wood02,sultan06,witten07}. More precisely, a log-normal distribution describes a fragmentation process in which all pieces are broken successively into two parts, such that any new fragment is further broken into two pieces where the breaking point is uniformly distributed along the fragment~\cite{sultan06,Villermaux03,Villermaux04}. In contrast, the Gamma distribution emerges from a fragmentation process where all the breaking points are uniformly distributed along the unbroken line, prior to the breaking that happens simultaneously for all points. Fig.~\ref{fig2}d shows that both distributions reasonably well describe the rapid decay of the tail of the distribution, but the Gamma distribution seriously overestimates the probability density at small lengths. 
 This originates from the fact that a hierarchical fragmentation process tends to generate less small fragments than a random one. A more rigorous test is done through the statistical $\chi^2$ test for goodness-of-fit, which confirms that the log-normal describes better the data. We checked that this
description is robust with respect to the chosen value of the threshold used in the segmentation procedure. 
The log-normal distribution accurately describes all the experimental data sets, so the crumpling is hierarchical rather than random. Earlier simulations~\cite{vliegenthart06} of crumpled sheets and experiments on unfolded sheets \cite{blair05} found a similar agreement with a log-normal distribution.

\begin{table}
  \caption{Results from the literature for the power-law variation of the force $F \propto (D/\Delta)^{\beta}$ for compactions of an $x$-dimensional object in $(x+1)$-dimensional space. The crumpliness exponent $\beta$ is  measured and  $\beta^\star$ is our theoretical prediction from the dimension $x^\star$ of the geometry of compaction. Topological constraints and the material properties may influence the value of the exponent. The first set corresponds to  a mixture of cases (a) and (b) of Fig.~\ref{fig0}; the second and third sets correspond to cases (b) and (c)  respectively.} 
  \begin{tabular}{|l|c|c|c|c|c|c|c|}
  \hline
Crumpled Object  and Ref.     &     $x$ & $x^\star$ &    $\beta$    & $\beta^\star$    \\ \hline
Paper       (this work)                & $2 $      & $1<x^\star<2$      & $1.3$       & $1<\beta^\star<2$ \\   
Mylar       \cite{matan02}                    & $2$      & $1<x^\star<2 $         & $1.89$       & $1<\beta^\star<2$ \\   
Tethered Membrane   \cite{Astrom04}                & $2$       & $x^\star\lesssim2$          & $1.85$     & $\beta^\star\lesssim 2$ \\  
\hline
Rods         \cite{boue07}                            & $1$         &  $1$        & $2$          & $2$         \\ 
Rods          \cite{Stoop08}          &  $1$         &  $1$        & $2.05$      & $2$         \\ 
\hline
Linearly Elastic Sheet  \cite{vliegenthart06}                   &  $2$         & $2$       & $4$        & $4$\\
Aluminum Foil   \cite{lin08}                  & $2$        & $2$       & $5.13$            & $6$   \\        
Phantom Sheet    \cite{vliegenthart06}                & $2$        & $2$        & $2.66$     & $2.5$ \\    
			\hline
		\end{tabular}
	\label{betas}
\end{table}

The conclusion is that folding and crumpling are very similar in nature and the crumpling process shall be viewed as arising from successive folding events. For ordered folding, simple models allow for predictions of the relations between force, compaction ratio and number of folds. Surprisingly, these are found to capture the main properties of crumpling also, in particular the hierarchical structure of the folds and the power law relation between the force and the compaction ratio. The analogy with folding then allows to define the 'crumpliness exponent' $\beta$ for various forms of crumpling process. 
Previous experiments and simulations in the literature have reported such exponents $\beta$ for the power-law dependence of the force on the compaction ratio. They can be explained using our arguments, i.e., solely by considering the dimensionality of the compaction process, the topological constraints and the mechanical properties of the material (e.g. ductility). Table~\ref{betas} summarizes various exponents $\beta$ from the literature, detailed below.

For the first set of data, Matan et al. \cite{matan02} used a compaction set-up similar to the one used here, and found an exponent of $\beta\approx 1.89$. The aspect ratio of their cylinder (height/diameter) is much smaller than ours; we thus anticipate that the  compaction is more $2d$ in nature, and hence expect a crumpliness exponent closer to $2$ than in our experiment.  The value of this exponent  can again be understood as a compaction process lying between 1d and 2d. As our arguments are based
on dimensionality, they allow to predict only bounds for
this type of experiments.
The simulations of \cite{Astrom04} on compacted tethered membranes found a value of $\beta\approx1.85$.  Except that loading is now biaxial, the  compaction process is in fact similiar to that of case (a) since the "height" fluctuations of the membrane are small. If the two directions were independent we recover $N\propto\phi^2$ and thus force $F=NF_0\propto \phi^2$. However, folding in one direction is inhibited by folding in the other direction. This effect will decrease the total number of folds leading to a crumpliness exponent $\beta^\star\lesssim2$.

For the second set, the analogy with case (b) is complete and both experimental~\cite{Stoop08}, theoretical~\cite{boue07} and numerical~\cite{Stoop08,daCunha09} results in the elastic regime are in agreement with the prediction of  a crumpliness exponent $\beta^\star=2$.  

The third set deals with experiments and simulations of 2d sheets crumpled inside 3d spheres.
The linearly elastic sheet \cite{vliegenthart06} is a perfect example of case (c) for which the crumpliness  exponent $\beta^\star=4$, in agreement with the simulations. The aluminum foil is  ductile with plastic deformations~\cite{lin08}  and the phantom sheet can cross itself~\cite{vliegenthart06}. These are more complicated cases; however we can estimate the exponent $\beta$. 
For the aluminum foil, one modifies the estimate of the elastic energy  $E_{el}$ of a folded sheet because of  ductility. Most of the folded sheet remains flat but the elastic energy is now concentrated in a region of length $D$ and width $1/\kappa_c$ with a curvature $\kappa_c$, which is a material constant: the curvature scale at which the material yields. The balance of the compaction energy and the  elasto-plastic energy then leads to $ F \sim B_n \kappa_c  \sim Eh_n^3 \kappa_c $. As $N=\phi$, one  finds $F(N) \sim N^3 F_0\sim F_0\phi^6$ leading to the value $\beta^\star= 6$, which is in fair agreement with experimental results \cite{lin08}.
For the phantom sheets, the absence of steric interactions implies that $F(N) \sim N F_0$ which is similar to the 1d case. The number of folded layers $N$ is then related to the compaction degree through $N \sim V/V_f$, where $V_f$ is the average volume occupied by the sheet. For high compactions, it is known that $V_f \sim R_g^{d_f}$ where $R_g$ is the radius of gyration and $d_f$ is the fractal dimension~\cite{Kantor87}. Thus we find $F \propto N \sim \phi^{\beta^\star}$, with $\beta^\star =d_f \simeq 2.5$~\cite{Kantor87,tallinen09}.

Finally, these arguments allow to explain why a wastebucket fills up so quickly when waste paper is crumpled into a ball. Using the equivalence between crumpling and folding, the wasted volume $\Delta V/V$ can be estimated from the folded case.  In the 3d case, this is given by
 \begin{equation}
 \frac{\Delta V}{V} \simeq \frac{\Delta^3-Nh\Delta^2}{\Delta^3}=1-\frac{h}{D}N ^{3/2}\, .
 \end{equation}
For $N=2^6$, $h=10^{-4}$m and $D=0.2$m, one has $\Delta V/V\approx 75\% $, which is an excellent estimate for the experimental observation that crumpling is a very inefficient compaction process. In conclusion, the observations presented here demonstrate a non-trivial relation between the force of compaction and the geometry of the crumpled configuration. A potential application of this result would be to invert this problem, and deduce the force  through analyzing cross sections of crumpled sheets. Since our arguments are generic, they should hold at the nano-scale and could provide a simple framework to understand crumpled graphene structures, such as graphene-based supercapacitors \cite{Graphene, Graphene2}.

\end{document}